\documentclass[useAMS,boldmath]{mn2e}
\usepackage{graphicx,lscape}
\usepackage{amssymb}

\newcommand{\mnras}{MNRAS}

\newcommand{\apjs}{ApJS}
\newcommand{\apjl}{ApJ}
\newcommand{\aj}{AJ}

\newcommand{\aap}{A\&A}

\newcommand{\degree}{\mbox{$^{\circ}$}}
\newcommand{\msun}{\mbox{M$_{\odot}$}}
\newcommand{\lsun}{\mbox{L$_{\odot}$}}
\newcommand{\mic}{\mbox{$\mu$m}}

\newcommand{\av}{A$_{\rm V}$}

\def \cm2{\mbox{cm$^{-2}$}}
\def \cm3{\mbox{cm$^{-3}$}}

\newcommand{\rf}{\par\noindent\hangindent 15pt {}}
\newcommand{\xten}[1]{\mbox{$\times 10^{#1}$}}

\newcommand{\gtappeq}{\raisebox{-0.6ex}{$\,\stackrel
{\raisebox{-.2ex}{$\textstyle >$}}{\sim}\,$}}

\title[Hi-GAL imaging of MYSOs]{Herschel Hi-GAL imaging of massive young stellar objects}

\author[F. A. Olguin {\it et al.}]{F. A. Olguin$^{1}$\thanks{E-mail:\,pyfao@leeds.ac.uk}, M. G. Hoare$^{1}$\thanks{E-mail:\,M.G.Hoare@leeds.ac.uk},
H. E. Wheelwright$^{1,2}$, S. J. Clay$^{1,3}$, W.-J. de Wit$^{1,4}$,
 \newauthor
I. Rafiq$^{1,5}$, S. Pezzuto$^{6}$, S. Molinari$^{6}$ \\
$^{1}$School of Physics \& Astronomy, E.C. Stoner Building, University
of Leeds, Leeds LS2 9JT, UK \\
$^{2}$MPIfR-MPG, Bonn, Germany \\
$^{3}$Astronomy Centre, Department of Physics and Astronomy, University of Sussex, Brighton, BN1 9QH, U.K.\\
$^{4}$European Southern Observatory, Alonso de Cordova 3107, Vitacura,
Santiago, Chile \\
$^{5}$Technische Universit\"at M\"unchen, Institut f\"ur Astronomische und Physikalische Geod\"asie
Arcisstrasse 21, 80333 M\"unchen, Germany \\
$^{6}$IAPS-Istituto di Astrofisica e Planetologia Spaziali, Via Fosso del Cavaliere 100, 00133, Roma, Italy. \\
}

\begin{document}

\date{}

\pagerange{\pageref{firstpage}--\pageref{lastpage}} \pubyear{2015}

\maketitle

\label{firstpage}

\begin{abstract}
We used Herschel Hi-GAL survey data to determine whether massive young stellar 
objects (MYSOs) are resolved at 70\mic\ and to study their envelope density 
distribution. Our analysis of three relatively isolated sources in the $l=30$\degree\ and $l=59$\degree\
Galactic fields show that the objects are partially resolved at
70\mic. The Herschel Hi-GAL survey data have a high
scan velocity which makes unresolved and partially resolved sources
appear elongated in the 70\mic\ images. We analysed the two scan
directions separately and examine the intensity profile perpendicular
to the scan direction. Spherically symmetric radiative transfer models
with a power law density distribution were used to study the
circumstellar matter distribution. Single dish sub-mm data were also
included to study how different spatial information affects the fitted
density distribution. The density distribution which best fits both
the 70\mic\ intensity profile and SED has an average index of
$\sim0.5$. This index is shallower than expected and is probably due
to the dust emission from bipolar outflow cavity walls not accounted
for in the spherical models.  We conclude that 2D axisymmetric
models and Herschel images at low scan speeds are needed to better
constrain the matter distribution around MYSOs.
\end{abstract}



\section{Introduction}

The formation of massive stars presents many challenges due to the
competing and interlinked roles of gravity, magnetic fields and
radiation. It is becoming clear through numerical simulations that 
material can continue to accrete on to a luminous, massive forming
star via an accretion disc despite the strong radiation pressure on
dust (Krumholz et al. 2009; Kuiper et al. 2010). Bipolar outflows
appear to be a ubiquitous ingredient in the star formation process 
driven by magnetic forces (Banerjee \& Pudritz 2007) which also helps
relieve the extreme radiation pressure (Cunningham et al. 2011). 

These competing infall and outflow processes shape the circumstellar
matter distribution around massive forming stars. The corollary of
this is that a detailed mapping of the circumstellar matter
distribution can be used to constrain models of massive star
formation. One way to probe the circumstellar matter close to the
protostar is to use the emission from the heated dust. This has some
advantages over using molecular line emission for which complex
chemical and excitation effects have to be taken into account before
the total gas density distribution can be recovered. There are also
disadvantages with using warm dust emission as it does not convey any
kinematic information and becomes optically thick for $\lambda < 100$\mic.
However, a full understanding of the dust emission also yields the temperature
distribution of the material which is an important input back into the
molecular line diagnostic process.

Different IR wavelengths will probe regions at different distances
from the central accreting source due to the temperature gradients. 
Typical temperature gradients vary from $T\propto r^{-3/4}$ in the
optically thick part to $T\propto r^{-1/2}$ in optically thin regions
(eg. Ivezi\'c \& Elitzur 1997). 
Taking the over-simplified approach of the Wien Displacement law to locate
where most of the emission arises from, we then see that the size
of the emitting region $r\propto \lambda^{4/3}$ or $r\propto
\lambda^{2}$ in optically thick and thin regions respectively. In the
immediate environment of a protostar we will have optically
thick conditions in the dense mid-plane regions whilst the bipolar
outflow cavities will be mostly optically thin. 

High resolution studies of the warm dust emission around massive
forming stars have included the 8-13\mic\ interferometric studies by
de Wit et al. (2007;2010;2011). These 40 milliarcsecond resolution
studies probe size scales of about 100 AU for typical distances of
nearby massive young stellar objects (MYSOs) which approaches the size 
of the dust sublimation radius of about 25 AU. The mid-IR visibilities 
are mostly matched by 2D axisymmetric radiative transfer models where 
most of the emission arises from the warm dust along cavity walls. A 
compact element inside the dust sublimation radius such as an accretion 
disc may be needed to explain the rising 8\mic\ visibilities. 

Since the dusty bipolar cavity walls are directly illuminated by the
central star then this is mostly optically thin emission. We would
therefore expect the size of the emitting region to scale as
$\lambda^{2}$. For single-dish observations then this size scale is
getting larger faster than the diffraction-limited resolution is
degrading. Hence, for a given diameter telescope it is better to use
the longest wavelength possible. For the thermal-IR regime from the
ground this is 24.5\mic\ at the far end of the Q band. Scaling from
the 60 milliarcsecond size for W33A at 13\mic\ (de Wit et al. 2007)
this would yield a size of 0.2\arcsec\ at 24.5\mic\ which would be
partially resolved by the 0.6\arcsec\ diffraction limit for 8 m
telescopes. This is what was found by de Wit et al. (2009) and
Wheelwright et al. (2012) who partially resolved a sample of massive
YSOs. de Wit et al. (2009) modelled the extended emission with
spherical models and concluded that the density distribution needed to
be $n\propto r^{-1}$. This was interpreted as requiring some
rotational support on these scales as it is shallower than the
$n\propto r^{-1.5}$ expected for free-falling material.  Wheelwright
et al. (2012) again used 2D axisymmetric radiative transfer models to
show that the 20\mic\ emission is also dominated by the warm envelope
dust along the cavity walls. This is seen explicitly in some edge-on
systems (e.g. De Buizer 2005; 2006).

The recent release of Herschel space telescope data
provides the highest resolution
images to date at far-IR wavelengths. If we extend the scaling above to
the shortest wavelength of Herschel of 70\mic\ then we would expect a
size of the emitting region of about 2\arcsec. This is comparable to
the 5\arcsec\ diffraction-limited beam of Herschel at 70\mic. Hence,
we expect to recover further spatial information on the somewhat
cooler dust located further from the central object, but not so far as
to be more influenced by the general molecular core environment and
ambient radiation field as is likely at the longer Herschel wavebands.

The most extensive observations of massive forming stars with Herschel
comes from the Hi-GAL survey of the Galactic Plane (Molinari et
al. 2010a) where most massive young stars are located. Here we examine
these data from the first two Science Demonstration Phase fields
observed as part of Hi-GAL (Molinari et al. 2010a) to determine
whether massive YSOs are indeed extended at 70\mic.

We define massive YSOs as deeply embedded, luminous
(L\gtappeq3000\lsun), mid-IR-bright, point sources that are not ionizing
their surroundings to form an ultra-compact H II region (UCHII). The
lifetime of this phase is about 10$^{5}$ years (Davies et al. 2011;
Mottram et al. 20111a). Davies et al. (2011) show that the luminosity
function of MYSOs and UCHII regions is consistent with the MYSOs
becoming swollen due to high accretion rates as predicted by the
models of Hosokawa \& Omukai (2009). This means their effective
temperatures are too low to ionise their surroundings until either
they stop accreting at high rates or grow to greater than about
30\msun, when they contract rapidly down to their main sequence radius. 

The sample of MYSOs we use comes from the Red MSX Survey (RMS)
(Lumsden et al. 2013). Starting from an initial colour selection of
mid-IR bright sources (Lumsden et al. 2002) from the MSX satellite
point source catalogue (Price et al. 2001) we have followed these up
with radio continuum observations to distinguish UCHII regions and
dusty planetary nebulae (Urquhart et al. 2009); determined kinematic
distances from $^{13}$CO observations and H I absorption (Urquhart et
al. 2008; 2012); and determined luminosities from far-IR fluxes
(Mottram et al. 2010; 2011b). All MYSOs discussed in this paper are
undetected at 5 GHz at the 1 mJy level.

In this paper we examine the Herschel imaging of the RMS MYSOs in two
lots of 2 square degree regions of the Galactic plane. The
peculiarities of the Hi-GAL 70\mic\ point spread function (PSF) are
discussed in section~\ref{psf} and the 70\mic\ imaging of the RMS
MYSOs in these regions is described in section~\ref{Observations}.
Radiative transfer modelling of the 70\mic\ emission is presented in
section~\ref{modelling} and the results discussed in section~\ref{results}. 
Conclusions are drawn in section
\ref{conclusions}.

\section{Hi-GAL 70\mic\ Point Spread Function}
\label{psf}
\begin{figure}
  \centering
  \includegraphics[width=8.5cm]{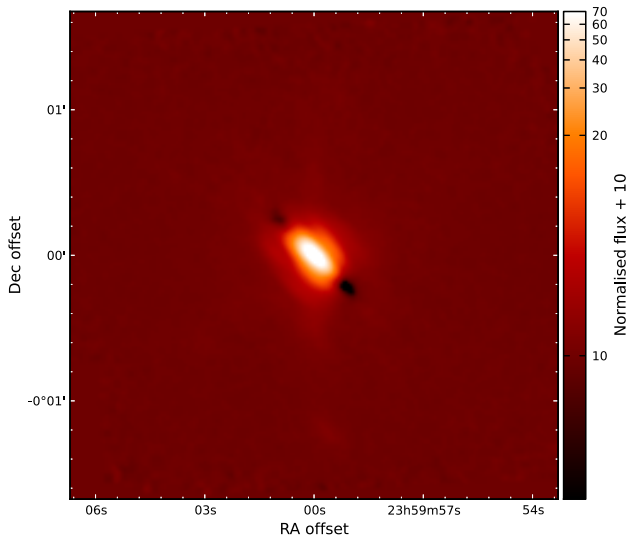} 
  \caption{Image of Vesta taken in parallel mode with a scan speed of 60\arcsec\ per second and
array-to-map inclination angle of $\alpha =$ 42.5\degree\ showing the details of the point spread
function. 
In order to use a logarithmic stretch, a constant value of 10 was added to the normalised fluxes.
}
  \label{fig:vesta}
\end{figure}

The Hi-GAL survey was carried out in a specially developed parallel
mode whereby the PACS and SPIRE instruments simultaneously scan the
sky at a fast rate of 60\arcsec\ per second (Molinari et
al. 2010). This causes the nominally circular PACS 70\mic\ beam with
FWHM of 5.3\arcsec\ to be smeared out in the scan direction with a
resultant size of 5.8\arcsec\ $\times$ 12.1\arcsec\ (Lutz 2012). A
high signal-to-noise representation of the PACS parallel mode
70\mic\ PSF is shown in figure~\ref{fig:vesta}, which shows an image of
the asteroid Vesta observed in similar conditions as
Hi-GAL data (see Lutz 2012). The first Airy ring can also
be seen smeared out along the scan direction (PA=42.5\degree) and a
dark spot is seen at one end of the scan direction.

The two Science Demonstration Phase (hereafter SDP) Hi-GAL fields were visually inspected to search for
objects that might be suitable PSF objects at 70\mic, i.e. bright,
unresolved and isolated. Such objects are rare and we only found two
in each of the SDP fields and their details are listed in table~\ref{tab:PSFstars}.
These appear to be all AGB or post-AGB stars and they were all located
away from the mid-plane consistent with an evolved, intermediate mass
population. Such stars are losing mass with dusty winds which makes
them suitable for IR PSF stars. However, as de Wit et al. (2009) found
these stars can also be extended and so have to be treated with
caution when using as PSF objects. 

\begin{table}
\caption{Parameters of the PSF objects found within the two Hi-GAL
  SDP fields at 70\mic.}
\label{tab:PSFstars}
\begin{tabular}{llccc}
\hline
Name & Nature & RA & Dec & $f_{70\mic}$ \\
& & \multicolumn{2}{c}{(J2000)} & (Jy) \\
\hline
V1362 Aql & Mira  & 18:48:41.9 & $-$02:50:28 & 66 \\
IRAS 18491$-$0207 & PAGB  & 18:51:46.2 & $-$02:04:12 & 80 \\
IRAS 19374+2359 & PAGB  & 19:39:35.5 & $+$24:06:27 & 29 \\
IRAS 19348+2229 & ? & 19:36:59.8 & $+$22:36:08 & 32 \\
\hline
\end{tabular}
\end{table}

\begin{figure}
  \centering
  \includegraphics[width=8.5cm]{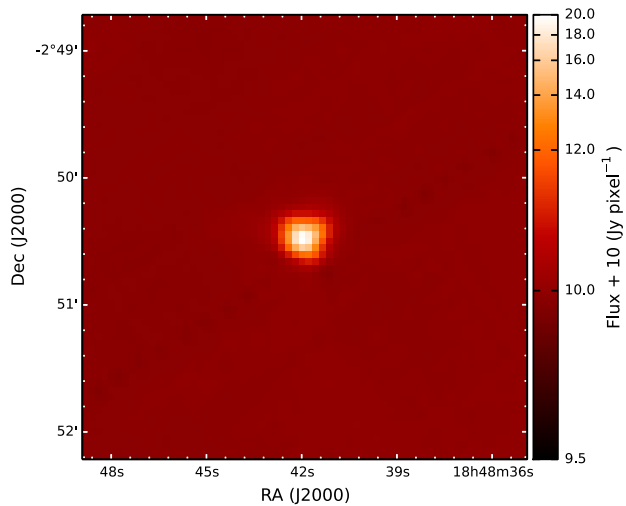}
  \caption{Image of the PSF star V1362 Aql in the naive map of the
    l=30 region. Note the offset cross shape caused by the addition of
  the nominal and orthogonal scans each of which has an elongated
  PSF.
  }
  \label{fig:PSFbin}
\end{figure}

An image of the PSF object V1362 Aql from the l=30\degree\ SDP field
is shown in figure~\ref{fig:PSFbin}. This is from the so-called naive
map constructed from the Hi-GAL data which adds together the
two different scan directions referred to as nominal and
orthogonal. Unfortunately this results in a complex PSF which is
basically a cross-shape. The naive maps we used had not had any
astrometric corrections applied, which also resulted in an offset
cross-shape. Such a complex PSF makes the search for extended emission
very difficult and certainly precludes the use of azimuthally averaged
radial profiles as we used for ground-based 24\mic\ imaging (de Wit et
al. 2009; Wheelwright et al. 2012). We decided to analyse images made from
the nominal and orthogonal scans separately which maximises the
resolution in the minor axis of the elongated PSF.

\begin{figure*}
  \centering
  \includegraphics[width=\textwidth]{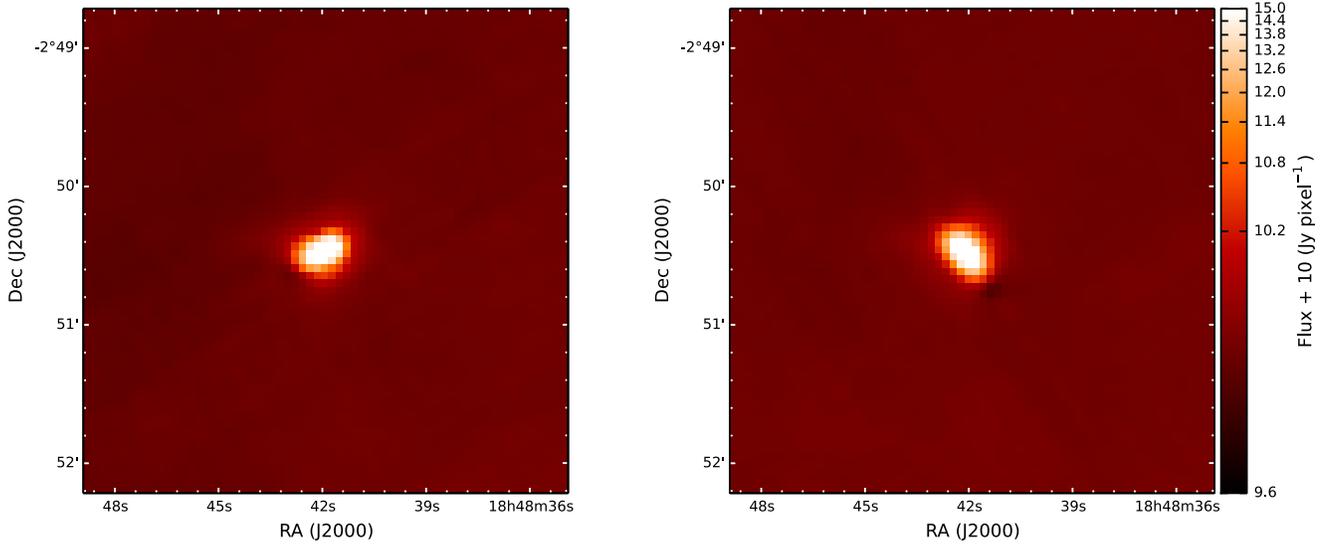}
  \caption{Nominal (left) and orthogonal (right) scan image of
    the PSF star V1362 Aql. 
    The flux scale is the same for both images.
  }
  \label{fig:psf1scans}
\end{figure*}

\begin{figure}[h!]
  \centering
  \includegraphics[width=8.5cm]{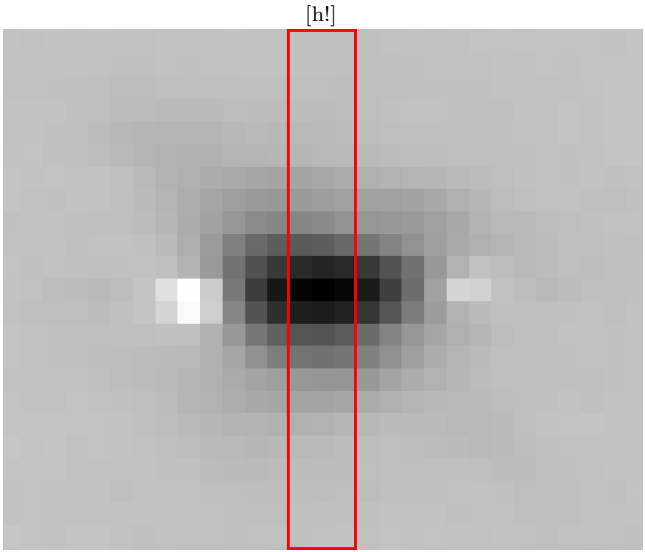}
  \caption{Image of Vesta ($\alpha = 42.5$\degree) rotated and rebinned to Hi-GAL resolution
    showing the slice along the minor axis used in this work (red box). Each pixel has a size
    of 3.2\arcsec\ and the image is $29\times23$ pixels in size.
    }
  \label{fig:vestaslice}
\end{figure}

We compared the Vesta PSF to that of
the PSF objects from the SDP field. This was to make sure that they
were consistent with each other 
as the PSF changes because the angle between the scan and
the detector axis (hereafter array-to-map angle, $\alpha$) changes with
map direction (Lutz 2012). The values of the array-to-map angle 
used are $\alpha=+42.5$\degree\ for the nominal direction, 
same as Vesta in figure \ref{fig:vesta}, and $\alpha=-42.5$\degree\ 
for the orthogonal direction (Molinari et al. 2010a).
Separate scan images of the PSF object V1362 Aql are shown in
figure~\ref{fig:psf1scans}. The structure in these images is similar
to that in the Vesta image in figure~\ref{fig:vesta}.
Each Vesta image was rotated so that the major axis
was horizontal and the scan direction pointing to the right, 
and then rebinned to the coarser pixel scale used in Hi-GAL as in 
figure~\ref{fig:vestaslice}. A vertical slice was then taken along the minor axis, 
three pixels wide and centred on the peak pixel. Each of the three columns 
were normalised to the central peak value and then a mean and standard 
deviation were taken from the three values at each offset position. 

The same procedure was applied to the nominal and orthogonal PSF
objects after first subtracting a mean background level determined in
an annulus surrounding the object, and the intensity profile of the
slices compared. The images were rotated so the positive scan
legs\footnote{For each map direction the telescope sweeps the sky in a
  pattern composed of several parallel scan legs, thus a leg can be
  either positive or negative depending on the direction with respect
  to the first scan leg.}  point in the same direction as the rebinned
Vesta image. To ensure the best comparison between objects at
the subpixel level, a 2D Gaussian was fitted to the image and the slice was shifted so that
the zero offset coincides with the Gaussian peak.

\begin{figure*}
  \centering
  \includegraphics[width=\textwidth]{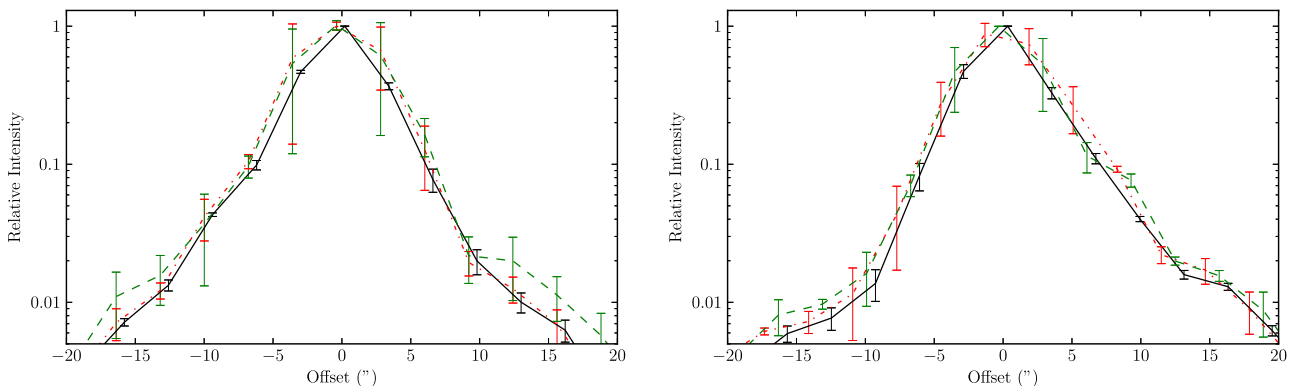}
  \caption{A comparison of the intensity profile of Vesta (solid black line) and PSF stars 
V1362 Aql (dash-dotted red line) and IRAS 18491$-$0207 (dashed green line) slices in 
the nominal (left) and orthogonal (right) map directions.
Note the agreement within the errors between Vesta and the PSF objects out to about the
1\% level.}
  \label{fig:psfcfvesta}
\end{figure*}

Figure~\ref{fig:psfcfvesta} shows that the intensity
slices of the rebinned Vesta and the PSF stars V1362 Aql and IRAS 18491-0207
drop to about 1\% of the peak or about 14\arcsec\ from the
centre, which is where the uncertainties on the background level of the
PSF objects become significant. 
Hence, since most of the MYSO targets are of similar or brighter flux than these PSF objects, 
comparing the MYSOs to these PSF objects would introduce more noise especially in the wings.
Therefore, we compared our
MYSOs to the much higher signal-to-noise image of Vesta from figure~\ref{fig:vesta}.

It is also worth noticing that the minor axis of the PSF objects is independent of whether
the object was present in one (e.g. IRAS 18491-0207 orthogonal) or both (e.g. IRAS 18491-0207 nominal) 
scan legs. However, for the purpose of modelling
the continuum emission (see Section \ref{modelling}) a Vesta image averaged with its reflection along
the major axis was used for sources present in both scan legs. 
The difference along the minor axis slice is less than 10\% between 
this averaged Vesta slices and the original one.

\section{Observations}
\label{Observations}
\subsection{70\mic\ Imaging of RMS MYSOs}
\label{Imaging}
\begin{figure*}
  \centering
  \includegraphics[width=\textwidth]{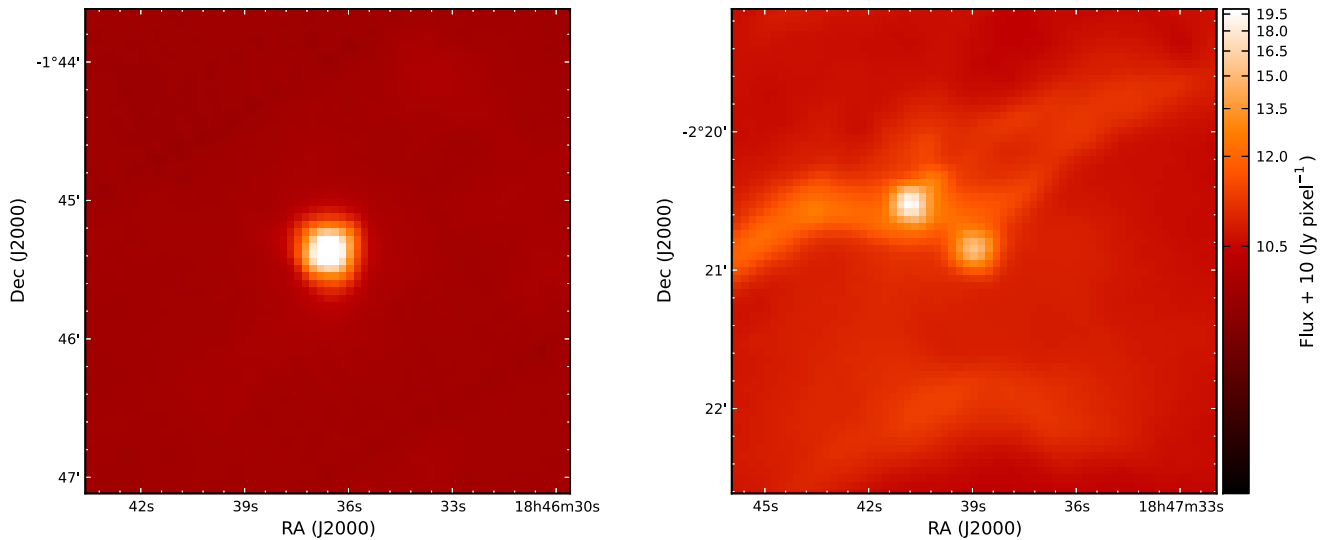}
  \caption{Image of the RMS MYSOs G030.8185+00.2729 (left) and G030.4117-00.2277 (right) from the naive map of the
    l=30 region.
    Note the partially resolved nature of G030.8185+00.2729 compared to the PSF star in figure~\ref{fig:PSFbin} and
    the background level and morphology compared to G030.4117-00.2277.}
  \label{fig:ysobin}
\end{figure*}

There are a total of 12 RMS MYSOs in the $l$=30\degree\ SDP field and 7 in
the $l$=59\degree\ field. We visually inspected each of these and found
that only one in the $l$=30\degree\ and two in the $l$=59\degree\ field were sufficiently isolated from
neighbouring sources and/or complex background (e.g. filamentary structures) emission to allow an
investigation of their extended emission. The parameters of the these MYSOs are given in table~\ref{tab:rmsmysos}.
Figure \ref{fig:ysobin} shows the naive 70\mic\ map of the brightest
of the MYSOs, G030.8185+00.2729, and an example MYSO ignored for its complex background
and the presence of a nearby object, G030.4117-00.2277. In our sampled objects, between 2 to 4 point sources 
were detected at 8\mic\ in the GLIMPSE/IRAC observations within 
the Herschel 70\mic\ resolution ($\sim6$\arcsec), but $\gtrsim 50$\% of the emission is dominated by the MYSO at 8\mic\ and
totally dominates in MIPS 24\mic\ images. Therefore, in what 
follows we will not consider multiplicity as major concern.

\begin{table*}
\caption{Parameters of the isolated RMS MYSOs found within the two Hi-GAL
  SDP fields at 70\mic.}
\label{tab:rmsmysos}
\begin{tabular}{lcccccccccc}
\hline
Name & RA & Dec & d & L & $\sqrt{\rm L}$/d $^{b}$& $f_{70\mic}$ & $f_{170\mic}$ & $f_{250\mic}$
& $f_{350\mic}$ & $f_{500\mic}$ \\
      &   &     & (kpc) & (\lsun) &(\lsun$^{1/2}$ kpc$^{-1}$) & (Jy) & (Jy) & (Jy) & (Jy) & (Jy) \\
\hline
G030.8185+00.2729 &  18:46:36.6 & $-$01:45:22 & 5.7 & 1.1\xten{4} & 18.4 & 321 & 269 & 131 & 57.6 & 22.7 \\
G058.7087+00.6607 & 19:38:36.8  & $+$23:05:43 & 4.4 & 4.4\xten{3} & 15.1 & 30.9 & 70.5 & 44.4 & 23.2 & 12.2 \\
G059.8329+00.6729 & 19:40:59.3  & $+$24:04:44 & 4.2 & 1.9\xten{3}$^{a}$ & 10.4 & 150 & 361 & 150 & 61.0 & 25.2 \\
\hline
\end{tabular}
\begin{minipage}{\textwidth}
$^{a}$ Note this object is in a cluster with several other YSOs within
about 5\arcsec. Its GLIMPSE 8\mic\ flux is only 20\% of the larger MSX
beam 8\mic\ flux and its total luminosity has therefore been reduced
by this amount to reflect the fact there may be other luminosity
sources in the large beam far-IR measurements of bolometric luminosity
(see Mottram et al. 2011).\\
$^{b}$ The physical size of a spherical dusty cloud heated to a particular temperature by a central 
source depends on the square root of the heating source luminosity which determines the spatial scales 
of the solution to the radiative transfer equation (Iveci\'c \& Elitzur 1997). 
The angular size is then inversely proportional to the distance. Therefore, it is an indicator
of how resolved is a source (see Section \ref{results}).
\end{minipage}
\end{table*}

\begin{figure*}
  \centering
  \includegraphics[width=\textwidth]{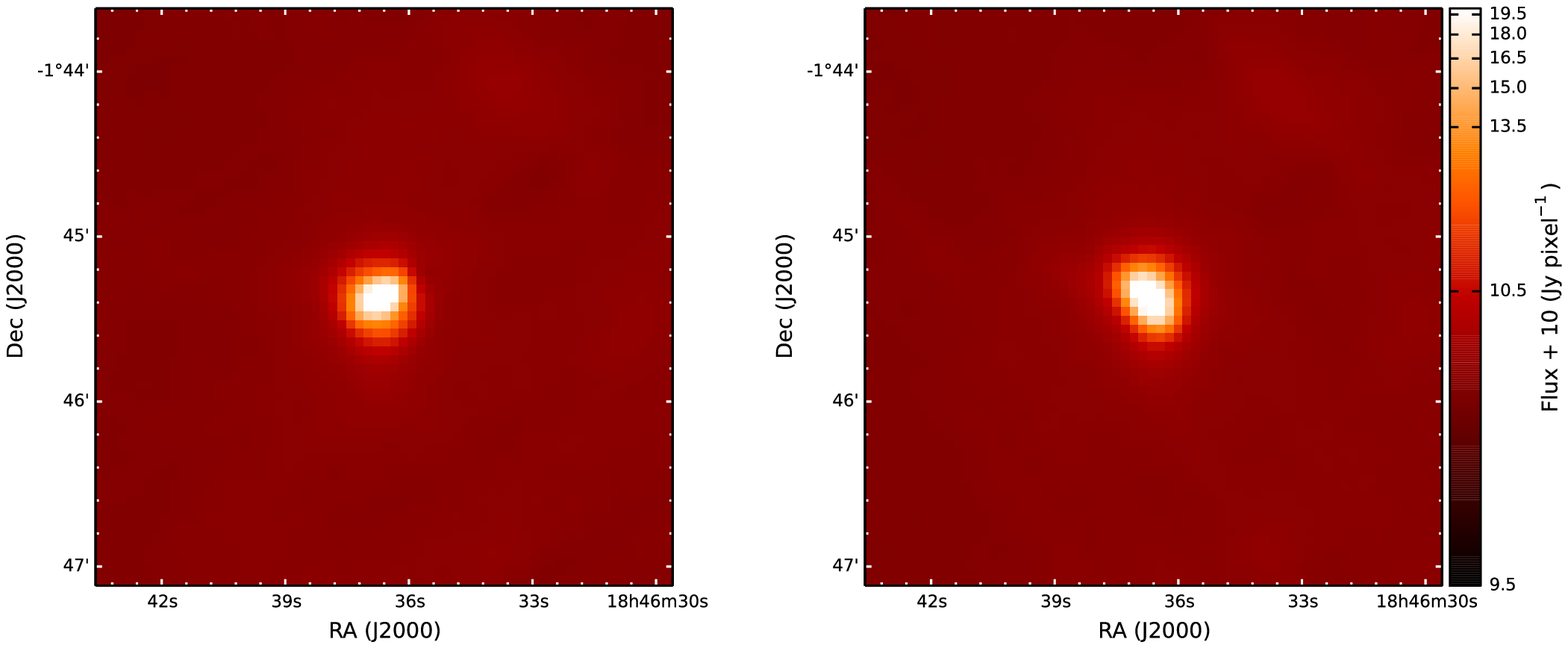}
  \caption{Nominal (left) and orthogonal (right) scan images for the
    RMS MYSO G030.8185+00.2729. 
    The flux scale is the same for both images.
  }
  \label{fig:yso2_images}
\end{figure*}

Comparing G030.8185+00.2729 in figure \ref{fig:ysobin} to the naive map of the
PSF star V1362 Aql in figure~\ref{fig:PSFbin}, clearly shows that the former
is more extended compared to the PSF star. Similarly, figure~\ref{fig:yso2_images} 
shows the nominal and orthogonal scan images for
G030.8185+00.2729 and these clearly show extended emission along
the minor axis direction compared to the PSF star in figure ~\ref{fig:psf1scans}. 
Intensity profiles for slices along the minor
axis were constructed for each of the MYSOs as described above 
for the PSF objects. In figures~\ref{fig:G030.8185_scan} to
\ref{fig:G059.8329_scan} these slices (square blue points with error
bars) are compared with those for Vesta (dashed cyan line). 
Again the 70\mic\ emission from the MYSO is clearly more
extended in the minor axis direction than the Herschel PSF for all the
MYSOs in both scan directions.

\subsection{Sub-mm Radial Profiles}

Although Herschel images of the same fields at 160, 250, 350 and 500\mic\ are available,
we limited the use of these images only for the SED. However, ground-based sub-mm
observations are available which are at higher resolution. Azimuthally averaged radial 
profiles at 870\mic\ were obtained from ATLASGAL 
images (Contreras et al. 2013) and at 450\mic\ from the SCUBA Legacy Catalogues 
(Di Francesco et al. 2008) and are shown in figure \ref{fig:longerwlg}. 
Annulii of 1.5 times the pixel size 
(6\arcsec\ at 870\mic\ and 3\arcsec\ at 450\mic)
were chosen to compute the average flux in each bin of angular 
distance from the peak. The errors in each bin were estimated
by the standard deviation of the fluxes in each annulus.

\section{Results}
\label{modelling}

To interpret the extended emission that we see at 70\mic\ we have
adapted the modelling procedure used by de Wit et al. (2009). We used
the same grid of spherical radiative transfer models for MYSOs that
was calculated by de Wit et al. (2009) using the DUSTY code (Ivezi\'c \&
Elitzur 1997). The grid of 120 000 models spans a range in density law
exponent $p$ where $n(r)\propto r^{-p}$ with $p$ varying from 0 to 2 in
steps of 0.5, A$_{\rm V}$ from 5 to 200 in steps of 5, and the ratio
of outer radius to sublimation radius, $Y \equiv R_{outer}/R_{sub}$, varying from 10 to
5000. For this study other model grid parameters were kept constant. 
These include the stellar effective temperature that was kept at 25 000 K	
corresponding to a B0 V star as the IR emission is insensitive to this
parameter. For the dust model we used the 'ISM' model as described in 
de Wit et al. (2009) that consists of Draine \& Lee (1984) graphite
and silicate with an MRN size distribution (Mathis et al. 1977).
This has a sub-mm emissivity law with a slope of $\beta = 2$.
The dust sublimation temperature was kept constant at 1500 K.

Each model was scaled to the appropriate luminosity and distance for
the MYSOs in table~\ref{tab:rmsmysos}.  A circular image of the
emergent 70\mic\ emission from the spherical model was generated and then
convolved with the Vesta PSF rebinned for the Hi-GAL pixel scale. 
As before, an intensity profile slice was generated from an average of the
three rows across the minor axis of the PSF direction normalised to
the peak pixel. These model slices were then compared to the observed
ones, both in the nominal and orthogonal scan directions.

\begin{figure*}
  \centering
  \includegraphics[width=\textwidth]{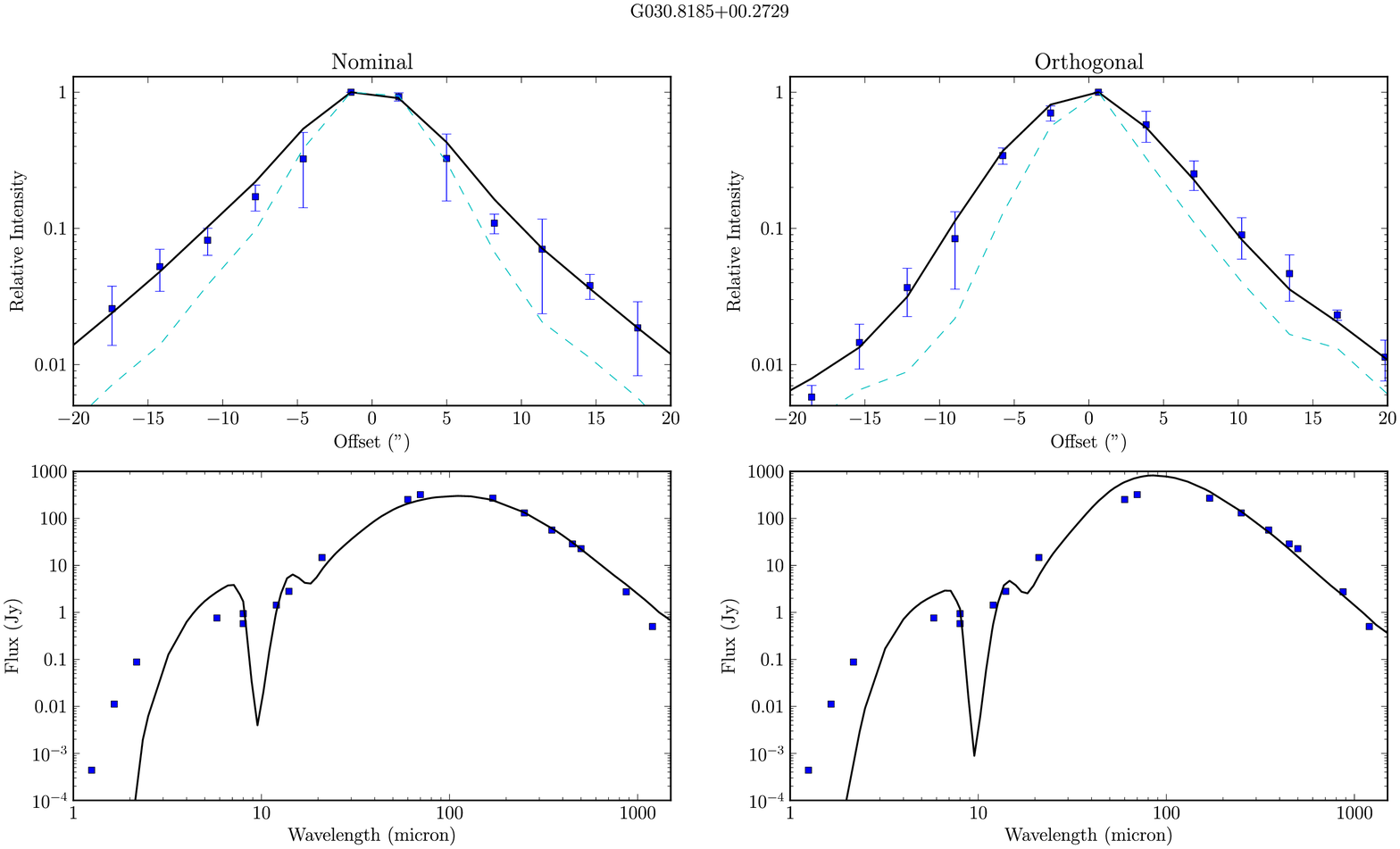}
  \caption{The combined best fit model (solid black) in terms of the
    70\mic\ scan profile (top) and SED (bottom) compared to the data (blue squares) for
    G030.8185+00.2729. The left hand panels are for the scan in the nominal
    direction whilst the right hand panel is for the scan in the
    orthogonal direction. The Vesta PSF scan is
    shown in the top panel (dashed cyan) to illustrate the extended nature of the
    MYSO emission. An error of 10\% of the total observed fluxes was considered in the observed SED.
  }
  \label{fig:G030.8185_scan}
\end{figure*}

\begin{figure*}
  \centering
  \includegraphics[width=\textwidth]{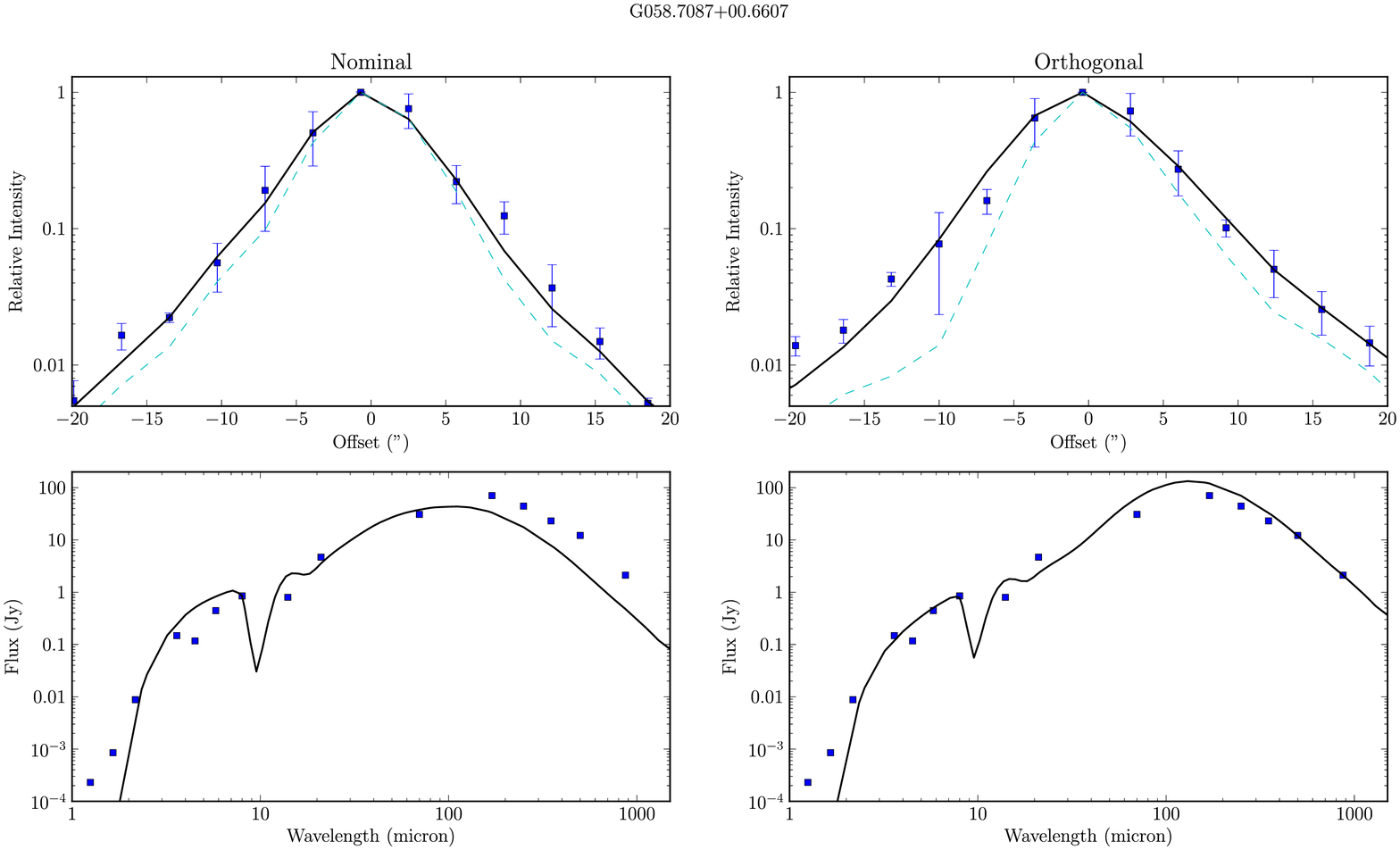}
  \caption{As in figure~\ref{fig:G030.8185_scan} but for G058.7087+00.6607.}
  \label{fig:G058.7087_scan}
\end{figure*}

\begin{figure*}
  \centering
  \includegraphics[width=\textwidth]{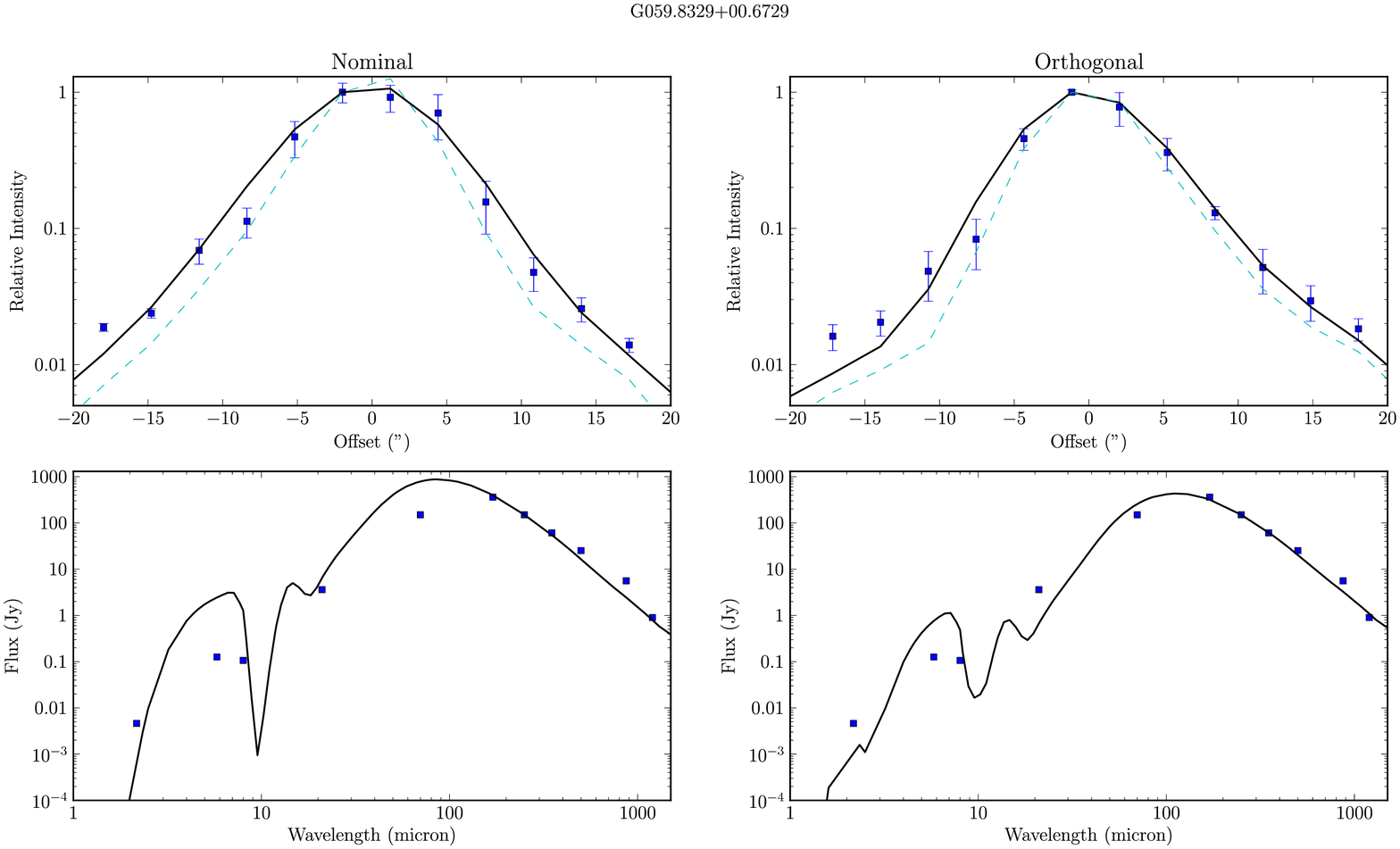}
  \caption{As in figure~\ref{fig:G030.8185_scan} but for G059.8329+00.6729.}
  \label{fig:G059.8329_scan}
\end{figure*}

Simultaneously with fitting the intensity profile slice we also fitted
the spectral energy distributions (SEDs). The luminosity used to scale
the models comes from fitting the SED. 
The data points in the SEDs
in figures~\ref{fig:G030.8185_scan} to \ref{fig:G059.8329_scan} are
from 2MASS (Skrutskie et al. 2006), GLIMPSE (Churchwell et al. 2009),
MSX (Price et al. 2001), Herschel (this work), sub-millimetre (Di
Francesco et al. 2008; Contreras et al. 2013) and millimetre 
observations (Beuther et al. 2002; Beltr\'an et al. 2006). 
Errors of 10\% were adopted for all the SED data
points to account for uncertainties in the absolute calibration across
different datasets. During the fitting procedure 
reduced-$\chi^2$ (hereafter $\chi^2$) values were
calculated for both the fits to the intensity slice and SED, where
the degrees of freedom were the number of fitted points minus one. These are
each placed in order of increasing $\chi^2$ and the model that is top
of the combined order is considered to be the best fitting model (e.g. de Wit 2009).

The best combined fits to the 70\mic\ intensity profile and SED for each direction are shown in
figures \ref{fig:G030.8185_scan} to \ref{fig:G059.8329_scan} whilst the 
parameters are listed in table \ref{tab:results}. 
In what follows, the results are not referred to any particular scan direction
unless otherwise stated.
Reasonable combined fits to the intensity profile of most of the objects are 
obtained with the models, with $\chi^2$ near 1 for the 70\mic\ profile. 
The SED fit shows that fluxes at $\lambda<3$\mic\ are always underestimated.
This is common for spherical models as they do not account for near-IR light
being scattered and escaping from the bipolar outflow cavities 
(de Wit et al. 2010).
The average power law index of the best fitting models is $p=0.5$.

\begin{figure*}
  \centering
  \includegraphics[width=\textwidth]{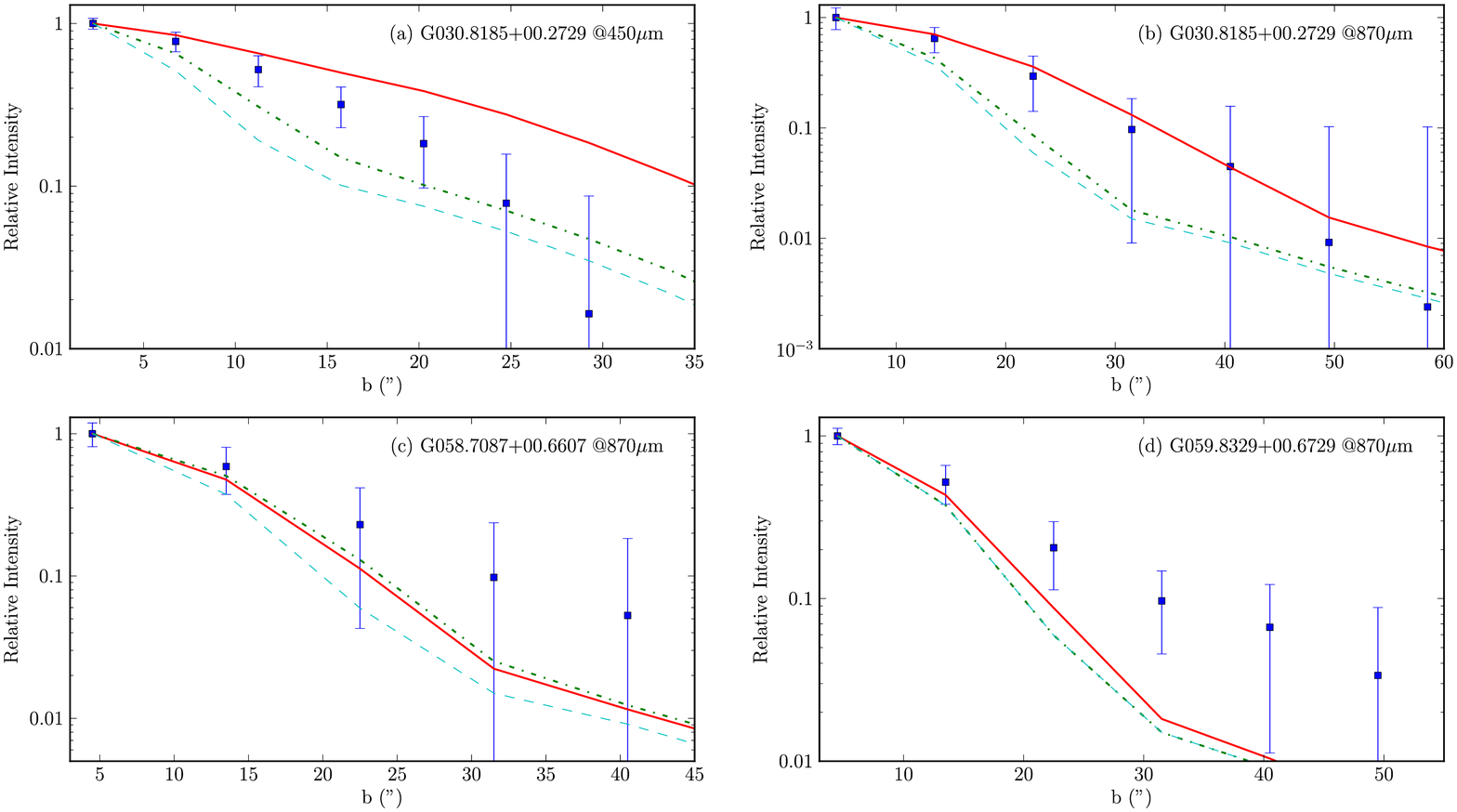}
  \caption{Azimuthally averaged radial profiles of sub-mm observations for each source (blue squares). The radial profiles from the best fitted models using the 
  70\mic\ intensity profile and SED for nominal (solid red line) and orthogonal (dash-dotted green line) directions are also shown. 
  \textit{(a)} and \textit{(b)} show G030.8185+00.2729 radial profiles for 450 and 870\mic, respectively.
  \textit{(c)} and \textit{(d)} show the 870\micron\ radial profile of G058.7087+00.6607 and G059.8329+00.6729, respectively.
  The dashed cyan line corresponds to the PSF.
  The impact parameter $b$ corresponds to the angular distance to the peak of the emission in the plane of the sky.
  }
  \label{fig:longerwlg}
\end{figure*}

Figure \ref{fig:longerwlg} show the profiles at 450 and 870\mic\ as
seen in the combined fit of the SED and 70\mic\ intensity profile.
Sub-mm radial profiles were also used instead of the 70\mic\ slices to analyse 
the effects of the spatial information from different wavelengths on the density 
distribution fit. These profiles constrain the density distribution of the cool outer 
regions. The combined 850/450\mic\ profile and SED best fits have an average exponent $p=0.5$, 
which is consistent with the combined 70\mic\ profile and SED fit.

Finally, fits to each individual observation were also calculated. The results
show that the best fits to radial profiles have exponents between 1 and 2,
whilst the fits to SEDs have exponents between 0 and 0.5. In addition, the exponent of
the best fitted models to the radial profiles is independent of \av\ and the size
of the cloud. 

\section{Discussion}
\label{results}

The average power law index is shallower than the $p\sim1$ exponent in the sample of 
de Wit et al. (2009) who fitted 24.5\mic\ intensity profiles and SED.
In addition, our averages do not agree in general with other works which have found that 
the values of the power law index vary between $1$ and $2$ 
by combining SED and sub-mm observations (e.g. Mueller et al. 2002).
Nevertheless, in the particular case of G030.8185+00.2729, Williams et al. (2005)
obtained an exponent of $0.5$ by using the SED and 850\mic\ radial profile, 
which agrees with our results.
The results of Williams et al. (2005) where obtained by including 
in the SED points with $\lambda\geqslant12$\mic\ whilst Mueller et al. (2002) included
points with $\lambda\geqslant30$\mic. We experimented with also only fitting data with 
$\lambda\geqslant30$\mic and found that in general values of exponents are 0.5 lower than using 
the whole SED, and therefore the exponents are still between 0 and 1.
We obviously do not expect to match the average results of Mueller et al. and de Wit et al.
since our sample has only 3 objects.

The power law indexes for the slice only and 870\mic\ only cases vary between
$1.5$ and $2$, whilst in the 450\mic\ only cases its value is $p=1$. 
These values are consistent with those obtained by 
Beuther et al. (2002), who found an average value of $p=1.6$, even though
they derived their values from a power law fitted to the radial profiles instead of 
doing the radiative transfer. Of course, these models do not fit the
SED well.

On the other hand, the power law indexes for the fits to the SED only
vary between $0$ and $0.5$. This is similar to previous studies that
use a dust emissivity law with a slope of $\beta=2$ (e.g. Chini et
al. 1986). As discussed by Hoare et al. (1991), a shallower dust
emissivity law allows fits with a steeper density distribution. In
fact, inspection of the SED fits at $\lambda>100$\mic\ in figure \ref{fig:G058.7087_scan} appears to show
that the $\lambda^{-2}$ emissivity law used in the modelling is
slightly too steep. Moreover, a study of the dust emissivity law
in these two regions by Paradis et al. (2010) shows that the emissivity
slope should be $\sim1.5$ in $l=30$\degree\ and $\sim1$ in $l=59$\degree\ 
for a dust temperature of 30 K.
Our higher value of the slope would also explain the large values of \av, for
steeper emissivity laws need more dust mass to match the dust emission in the 
far-IR/submm.

The values of \av\ range between 95 and 200 for the combined SED and slice fit, but
most of the sources have an \av\ of 200.
This is consistent with them being massive, young and embedded objects in their 
parental clouds. 
However, de Wit et al. (2009) found lower values than ours.
In the particular case of G030.8185+00.2729, Williams et al. (2005) found a value
$\sim 4$ times larger than ours, and using the method of Mueller et al. (2002)
we obtain similar values of \av\ as those obtained by considering all the points in the SED. 
Either way, all these results point towards values of \av\ greater than 90 magnitudes, and
the inclusion of points at smaller wavelengths does not determine the value of \av\ though
it helps to constrain it. The value of \av\ seems to be determined by the amount of dust 
necessary to reproduce the far-IR/sub-mm dust emission given its emissivity law.

\begin{table*}
\caption{Best fit model parameters for the fits to the 70\mic\ intensity slice and the 450\mic\ and 870\mic\ radial profiles.}
\label{tab:results}
\begin{tabular}{lllccccccc}
\hline
Name 	&	Scan  	&	 Fit	&	  $p$ 	&	 A$_{\rm V}$ 	&	 $Y^{a}$  	&	 $\chi^2_{SED}$ 	&	 $\chi^2_{70\mic}$	
&	$\chi^2_{450\mic}$	&	$\chi^2_{870\mic}$	\\
\hline																			
G030.8185+00.2729	&	 nominal	&	SED + 70\mic\	&	1.0	&	200	&	5000	&	109	&	1.4	&	4.7	&	0.23	\\
			&			&	SED + 450\mic\	&	0.5	&	170	&	2000	&	40	&	3.2	&	0.4	&	1.48	\\
			&			&	SED + 870\mic\	&	0.5	&	120	&	5000	&	73	&	3.6	&	4.1	&	0.04	\\
			&	 orthogonal	&	SED + 70\mic\	&	0.5	&	200	&	1000	&	77	&	0.7	&	1.9	&	2.17	\\
			&			&	SED + 450\mic\	&	0.5	&	170	&	2000	&	40	&	1.6	&	0.4	&	1.48	\\
			&			&	SED + 870\mic\	&	0.5	&	120	&	5000	&	73	&	13.6	&	4.1	&	0.04	\\
\vspace{1.5 mm}\\
																				
G058.7087+00.6607	&	 nominal	&	SED + 70\mic\	&	1.0	&	120	&	5000	&	132	&	0.6	&	\ldots	&	0.48	\\
			&			&	SED + 870\mic\	&	0.5	&	150	&	5000	&	62	&	3.2	&	\ldots	&	0.25	\\
			&	 orthogonal	&	SED + 70\mic\	&	0.0	&	95	&	5000	&	45	&	2.6	&	\ldots	&	0.35	\\
			&			&	SED + 870\mic\	&	0.5	&	150	&	5000	&	62	&	2.8	&	\ldots	&	0.25	\\
\vspace{1.5 mm}\\
																					
G059.8329+00.6729	&	 nominal	&	SED + 70\mic\	&	0.5	&	200	&	1000	&	4800	&	4.8	&	\ldots	&	1.42	\\
			&			&	SED + 870\mic\	&	0.5	&	200	&	5000	&	101	&	6.2	&	\ldots	&	0.57	\\
			&	 orthogonal	&	SED + 70\mic\	&	0.0	&	200	&	2000	&	403	&	1.5	&	\ldots	&	1.88	\\
			&			&	SED + 870\mic\	&	0.5	&	200	&	5000	&	101	&	2.0	&	\ldots	&	0.57	\\
\hline
\end{tabular}
\begin{minipage}{\textwidth}
 \textbf{Notes: } The values of $\chi^2$ correspond to the reduced $\chi^2$. \\
$^{a}$ $Y \equiv R_{outer}/R_{sub}$ with $R_{outer}$ the outer radius and $R_{sub}$ the sublimation radius of the envelope\\
\end{minipage}
\end{table*}

Table \ref{tab:rmsmysos} shows the values of the $\sqrt{\rm L}$/d
ratio, which has previously found to be a good indicator of how resolved these
objects are (e.g. Wheelwright et al. 2011). This ratio is 
proportional to the angular size of the inner rim of the spherical envelope 
(Iveci\'c \& Elitzur 1997) and, as is shown in table \ref{tab:results}, 
the envelopes have sizes a few thousands times the sublimation radius 
($\sim25$ au for $\rm L =10^4$\lsun), and should therefore be resolved at longer wavelengths.
Figures 8 to 10 show the degree to which the objects are resolved at 70\mic\ agrees with
this. 

\begin{figure*}
  \centering
  \includegraphics[width=8.5 cm]{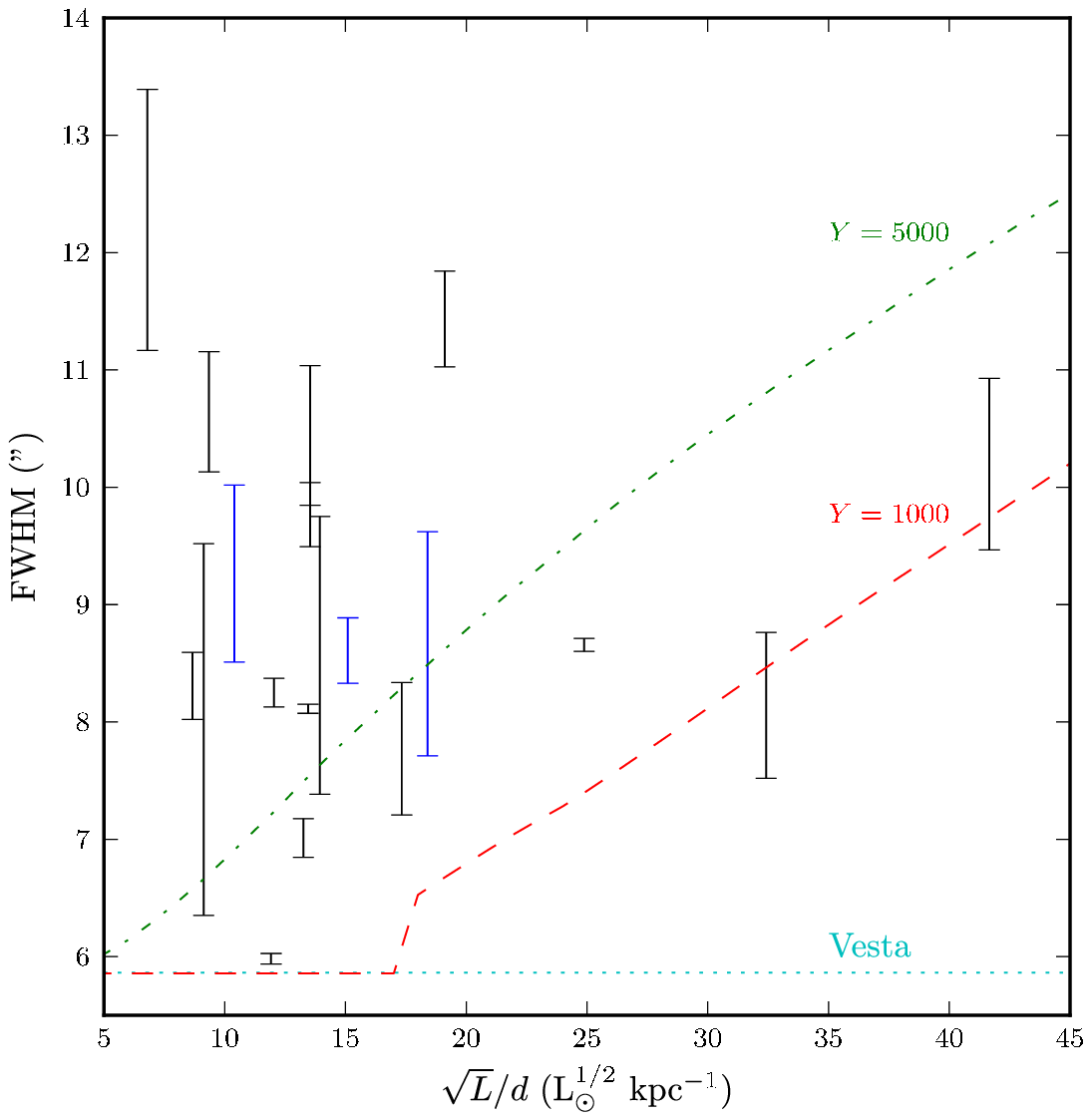}
  \caption{Relation between $\sqrt{\rm L}/\rm d$ and FWHM of a 1D Gaussian fitted to the 70\mic\ slices
  from MYSOs in the $l=30$\degree\ and $l=59$\degree\ fields and models. 
  The bar ranges are defined by the FWHM of the fit to the nominal and orthogonal directions, and the blue bars
  correspond to our sample.
  The predicted relation from two models with $p=0.5$ and A$_{\rm V}=200$ is shown in red dashed line for $Y=1000$ and in greed dot-dashed line for $Y=5000$.
  Model images were convolved with the nominal Vesta PSF and a mean error of 1.2\arcsec\ is estimated for the Gaussian fit.
  The horizontal cyan dotted line represents the FWHM of the Vesta nominal slice.
  }
  \label{fig:radvld}
\end{figure*}

To explore whether this extends to the other 16 MYSOs with more complex background/neighbouring sources,
we repeat the procedures used in the previous sections to obtain slices
from the other sources in the Hi-GAL fields and from two models with similar physical
properties as those obtained by the radiative transfer results, and then
fitted 1D Gaussian to measure the FWHM of these slices to see how resolved
the sources are. Figure \ref{fig:radvld} shows the relation between the FWHM and
the $\sqrt{\rm L}/\rm d$ ratio. 
All observed sources are consistent with the models with some of them more
extended due to the complex background.

\section{Conclusions}
\label{conclusions}

We have presented 70\mic\ observations made with the Herschel PACS
instrument towards two regions of the Galactic plane and identified
three relatively isolated MYSOs. The peculiarities of the Hi-GAL
survey PSF and its effects on the MYSOs observations were analysed.
The sources in our sample are all partially resolved at
70\mic.

Using spherical radiative transfer models to simultaneously fit
the 70\mic\ profile and SED, we find we need a density law 
exponent of around 0.5. This is shallower than we previously
found from fitting partially resolved 24.5\mic\ ground-based imaging,
though both observations give an exponent between 0 and 1.
It is also shallower than expected for infalling material 
($p=1.5$). This could be due to rotational support, but since
the emitting region is well outside of the expected disk/centrifugal 
radius (less than a few thousands AU, e.g. Zhang 2005)
this is unlikely. It is more likely due to warm dust
along the outflow cavity walls as seen in the mid-IR.
We will investigate this further using 2D axisymmetric models.
Intrinsic asymmetry could explain why we do not always get the same
results on the same object from the nominal and orthogonal scan directions.

Finally, the images at 70\mic\ were smeared along the scan direction
due to the scan speed. Moreover, the lack of PSF stars in the
fields does not allow a characterisation of the PSF specific for these observations.
Therefore, slow scan data, with a better behaved PSF, will be better to map and 
constrain the matter distribution of MYSOs. In particular, if the dust emission at 70\mic\ 
comes from a non-spherical structure like bipolar cavity walls, data
at 70\mic\ can provide useful insights for 2D models.

\section*{References}
\vspace{-10pt}
\rf{}
\rf{Banerjee, R. \& Pudritz, R. E. 2007, ApJ, 660, 479}
\rf{Beltr\'an, M. T., Brand, J., Cesaroni, R., Fontani, F., Pezzuto, S., Testi, L., \& Molinari, S. 2006, A\&A, 447, 221}
\rf{Beuther, H., Schilke, P., Menten, K. M., Motte, F.; Sridharan,  T. K.; Wyrowski, F. 2002, ApJ, 566, 945}
\rf{Chini, R., Kruegel, E., \& Kreysa, E.\ 1986, \aap, 167, 315}
\rf{Churchwell, E., et al. 2009, PASP, 121, 213}
\rf{Contreras, Y., Schuller, F., Urquhart, J.~S., et al.\ 2013, \aap, 549, A45}
\rf{Cunningham, A. J., Klein, R. I., Krumholz, M. R., McKee, C. F. 2011, ApJ, 740, 107}
\rf{Davies, B. et al. 2011, MNRAS, 416, 972}
\rf{De Buizer, J. M., 2006, ApJ, 642, L57}
\rf{De Buizer, J. M., 2007, ApJ, 654, L147}
\rf{de Wit, W.~J., Hoare, M.~G., Oudmaijer, R.~D., \& Mottram, J.~C.\ 2007, \apjl, 671, L169}
\rf{de Wit, W.~J., Hoare, M.~G., Fujiyoshi, T., et al.\ 2009, \aap, 494, 157}
\rf{de Wit, W.~J., Hoare, M.~G., Oudmaijer, R.~D., \& Lumsden, S.~L.\ 2010, \aap, 515, A45}
\rf{de Wit, W.~J., Hoare, M.~G., Oudmaijer, R.~D., et al.\ 2011, \aap, 526, L5 }
\rf{Di Francesco, J., Johnstone, D., Kirk, H., MacKenzie, T., \& Ledwosinska, E., 2008, ApJS, 175, 277}
\rf{Draine, B. T., \& Lee, H. M. 1984, ApJ, 285, 89}
\rf{Hoare, M.~G., Roche, P.~F., \& Glencross, W.~M.\ 1991, \mnras, 251, 584}
\rf{Hosokawa T. \& Omukai K., 2009, ApJ, 691, 823}
\rf{Ivezic, Z., \& Elitzur, M. 1997, MNRAS, 287, 799}
\rf{Krumholz, M. et al. 2009, Sci 323, 754}
\rf{Kuiper, R., Klahr, H., Beuther, H., Henning, T., 2010, ApJ, 722, 1556}
\rf{Lumsden, S.~L., Hoare, M.~G., Oudmaijer, R.~D., \& Richards, D.\ 2002, \mnras, 336, 621}
\rf{Lumsden, S.~L., Hoare, M.~G., Urquhart, J.~S., et al.\ 2013, \apjs, 208, 11}
\rf{Lutz, D. 2012, Herschel internal document PICC-ME-TN-033}
\rf{Mathis, J. S., Rumpl, W., \& Nordsieck, K. H. 1977, ApJ, 217, 425}
\rf{Molinari, S., et al., 2010a, PASP, 122, 314}
\rf{Molinari, S., et al., 2010b, A\&A, 518, L100}
\rf{Mottram, J. C., et al. 2010, A\&A, 510, 89}
\rf{Mottram, J. C., et al. 2011, A\&A, 525, 149}
\rf{Mottram, J. C., et al. 2011, ApJ, 730, L33}
\rf{Mueller, K.~E., Shirley, Y.~L., Evans, N.~J., II, \& Jacobson, H.~R.\ 2002, \apjs, 143, 469}
\rf{Paradis, D., Veneziani, M., Noriega-Crespo, A., et al.\ 2010, \aap, 520, L8}
\rf{Price S. D., et al., 2001, \aj, 121, 2819}
\rf{Skrutskie, M. F. et al. 2006, AJ, 131, 1163}
\rf{Urquhart, J. S., et al. 2008, A\&A, 487, 253}
\rf{Urquhart, J. S., et al. 2009, A\&A, 501, 539}
\rf{Urquhart, J. S., et al. 2012, MNRAS, 420, 1656}
\rf{Wheelwright, H.~E., de Wit, W.~J., Oudmaijer, R.~D., et al.\ 2012, \aap, 540, A89}
\rf{Williams, S.~J., Fuller, G.~A., \& Sridharan, T.~K.\ 2005, \aap, 434, 257}
\rf{Zhang, Q.\ 2005, Massive Star Birth: A Crossroads of Astrophysics, 227, 135}

\label{lastpage}

\end{document}